\documentclass[12pt,showpacs,preprintnumbers,amssymb]{revtex4}
\usepackage{graphicx}
\usepackage{dcolumn}
\usepackage{bm}

\begin{document}
\title{Origin of the Transition Inside the Desynchronized State in Coupled
  Chaotic Oscillators}

\author{Chil-Min Kim}
\author{Won-Ho Kye}
\author{Sunghwan Rim}
\affiliation{National Creative Research Initiative Center for  Controlling
  Optical Chaos\\ Pai-Chai University, Taejeon 302-735, Korea}

\author{Dong-Uk Hwang}
\author{Inbo Kim}
\author{Young-Jai Park}
\affiliation{Department of Physics and Basic Research Institute, Sogang
  University, Seoul 100-611, Korea}

\author{Eok-Kyun Lee}
\affiliation{Department of Chemistry and School of Molecular Science
  (BK21), Korea Advanced Institute of Science and Technology, Taejon 305-701,
  Korea}

\begin{abstract}
 We investigate the origin of the transition inside the desynchronization state via
 phase jumps in coupled chaotic oscillators. We claim that the transition
 is governed by type-I intermittency in the presence of noise whose
 characteristic relation is
 $\langle l \rangle \propto \exp(\alpha|\epsilon_t -\epsilon|^{3/2})$
 for $\epsilon_t -\epsilon <0$ and
 $\langle l \rangle \propto (\epsilon_t -\epsilon)^{-1/2}$ for
 $\epsilon_t -\epsilon >0$, where $\langle l \rangle$ is the average length of
 the phase locking state and $\epsilon$ is the coupling strength. To justify our claim we obtain analytically the tangent point,
 the bifurcation point, and the return map which agree well with
 those of the numerical simulations. 
\end{abstract}


\maketitle

\section{Introduction}
 Recently synchronization phenomena in coupled chaotic oscillators have been
 extensively studied because of their fundamental importance in areas of
 science and technology such as laser dynamics, electronic circuits, and
 biological systems\cite{Apps,Tut,Strogatz}. Due to interaction between two coupled
 chaotic oscillators, various features of synchronization are observed
 depending on the coupling strength. For example, when two identical chaotic
 oscillators are coupled, they can be synchronized either perfectly\cite{Sync}
 or intermittently\cite{Rim}. In another case, where they are coupled with
 slight parameter mismatch, non-synchronization, phase
 synchronization\cite{PhaseSync1,PhaseSync2,Rosa}, or lag
 synchronization\cite{Lag} is observed depending on the coupling strength.
 Among these features, the noteworthy one of our study is {\it phase
 synchronization} (PS)\cite{PhaseSync1,PhaseSync2,Rosa}. 
 As is widely understood, above the critical
 strength of the coupling for the transition to PS, suitably defined phases of
 two chaotic oscillators are locked while their amplitudes remain chaotic and
 uncorrelated. And below the critical value, the phases of two oscillators are
 intermittently unlocked, that is, $2\pi$ phase jumps interrupt the phase
 locking states irregularly. So it can be said that 
 the transition from nonsynchronous
 state to PS state is typically accompanied by an intermittent sequence of $2\pi$
 phase jumps\cite{PhaseSync1,Lee}.

 To explain the transition mechanism to PS, two different approaches have
 been introduced: topological\cite{PhaseSync1,PhaseSync2,Rosa} and
 statistical\cite{Lee,IBKim}. The topological approach assumes that the
 behavior of coupled chaotic oscillators is analogous to that of a chaotic
 oscillator driven by an external chaotic signal\cite{Topo}. From this study
 it was concluded that the phase jump phenomenon stems from a boundary
 crisis\cite{FBasin} mediated by an unstable-unstable pair bifurcation, which
 is termed eyelet intermittency\cite{Eyelet}. Meanwhile, the statistical approach focuses on
 a phase equation which describes the phase difference of coupled chaotic
 oscillators. A potential modulated
 by multiplicative noise was derived by the analysis of the phase equation 
 and it was concluded that the transition
 mechanism is similar to that of eyelet intermittency\cite{Lee}.

  By taking the statistical approach, however, we are to show in this paper that the
 characteristic relation of the average length of the phase locking state (or average laminar length) 
 on the coupling strength before the transition to PS in coupled R\"{o}ssler
 oscillators follows the relation of type-I intermittency in the presence of
 noise. To validate our argument, we derive a one-dimensional phase equation
 from the coupled R\"{o}ssler oscillators and obtain the bifurcation
 point, the tangent point that is a $\pi/2$ phase locking state, and the return
 map analytically, which exactly agree with those of the numerical simulations. 
 And then we obtain the characteristic relations of
 $\langle l \rangle \sim \exp(\alpha |\epsilon_t -\epsilon|^{3/2})$ for
 $(\epsilon_t -\epsilon) <0$ and
 of $\langle l \rangle \sim (\epsilon_t -\epsilon)^{-1/2}$ for
 $(\epsilon_t -\epsilon) >0$ which are the same as those of type-I
 intermittency in the presence of noise\cite{Kye,Eckman}, where $\langle l \rangle$ 
 is the average length of the phase locking state, $\epsilon$ is
 the coupling strength and $\epsilon_t$ is the tangent bifurcation point.

In our approach, the analytic scaling rule is obtained from the Fokker-Planck equation \cite{FPE} 
and so it describes the asymptotic behavior near PS regime in which the
effective noise can be reasonably approximated to be Gaussian \cite{IBKim,Kye}.  
For this reason, it seems that critical coupling does not appear in our formalism.
However in real systems there is one critical coupling $\epsilon_c$  which is defined
as a border of existence of phase slips, because of a finite amplitude of the effective noise.

In section II, we will obtain phase difference equation of coupled R\"ossler oscillators
 from one-dimensional phase equation and compare the analytical
 and numerical return maps of the phase difference in Section III. After that we will
 obtain the characteristic relations of intermittent phase locking time
 according to the coupling strength by Fokker-Planck equation in Section IV,
 discuss the results in Section V, and conclude our study in Section VI.

\section{Analytic study of Coupled R\"{o}ssler Oscillators}

  The transition to PS was first observed in coupled R\"{o}ssler oscillators
 which have a slight parameter mismatch \cite{PhaseSync1},
\begin{eqnarray}
        \dot{x}_{1,2}&=&-\omega_{1,2} y_{1,2} -z_{1,2} +\epsilon(x_{2,1}-x_{1,2}),\nonumber\\
        \dot{y}_{1,2}&=& \omega_{1,2} x_{1,2} +0.15 y_{1,2},\\
        \dot{z}_{1,2}&=& 0.2 +z_{1.2}(x_{1,2} -10.0), \nonumber
\end{eqnarray}
 where $\omega_{1,2}$ $(=1.0 \pm 0.015)$ are the overall frequency of each
 oscillator and $\epsilon$ is the coupling strength. By transforming the above
 equation to the polar form, we obtain the following one-dimensional phase
 equation, which describes the phase difference of the two oscillators:
\begin{eqnarray}
    \frac{d \phi}{dt}\!=\!\Delta \omega +A(\theta_1, \theta_2, \epsilon) \sin\phi
           + B(\theta_1, \theta_2)\epsilon+ \xi(\theta_1, \theta_2),
\end{eqnarray}
 where,
\begin{eqnarray}
        A(\theta_1, \theta_2, \epsilon) &=& (\epsilon+0.15)\cos(\theta_1+\theta_2)-\frac{\epsilon}{2}(\frac{R_2}{R_1}+\frac{R_1}{R_2}), \nonumber\\
        B(\theta_1, \theta_2) &=& -\frac{1}{2}(\frac{R_2}{R_1}-\frac{R_1}{R_2})\sin(\theta_1+\theta_2),\nonumber\\
        \xi(\theta_1, \theta_2) &=& \frac{z_1}{R_1}\sin(\theta_1)-\frac{z_2}{R_2}\sin(\theta_2),
\end{eqnarray}
 $\Delta\omega = \omega_1 - \omega_2$, $\phi =\theta_1-\theta_2$,
 $\theta_{1,2}=\arctan(y_{1,2}/x_{1,2})$, and
 $R_{1,2}=\sqrt{x_{1,2}^2 +y_{1,2}^2}$. Here $B$, $\xi$, and
 $(\epsilon + 0.15)\cos(\theta_1+\theta_2)$ in $A$ are fast
 fluctuating terms in comparison with slowly varying $\phi$.

\begin{figure}
\begin{center}
\rotatebox[origin=c]{0}{\includegraphics[width=8.5cm]{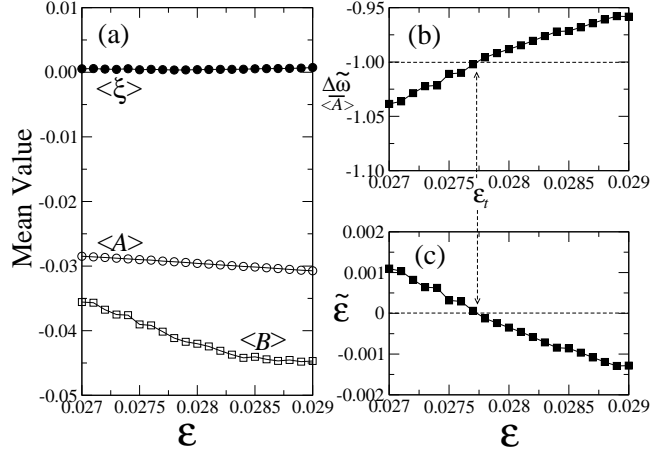}}
\caption{The values of $\langle A\rangle$, $\langle B \rangle$,
  $\langle \xi \rangle$, $\frac{\Delta \tilde{\omega}}{\langle A \rangle}$,
  and $\tilde{\epsilon}$ according to the coupling strength: (a) the mean
  values of A(circle), B(square), and $\xi$(filled circle); (b) and (c) the values of
  $\frac{\Delta \tilde{\omega}}{\langle A \rangle}$ and $\tilde{\epsilon}$
  (the arrows point out the bifurcation point $\epsilon_t$).}
\end{center}
\end{figure}

 Whereas, in the previous studies, the authors neglected all of the fast
 fluctuating terms\cite{Lee}, we found that they play a crucial role in the
 analysis of the transition mechanism.  Equation (2) can be transformed into
 the following simple form: 
\begin{equation}
        \frac{d \phi}{dt}=\Delta\tilde{\omega} +\langle A \rangle\sin\phi +\tilde{\xi},
\end{equation} 
  where $\tilde{\xi}=(\xi-\langle\xi \rangle)+(B-\langle B \rangle)\epsilon + (A-\langle A \rangle) \sin\phi$, and
        $\Delta\tilde{\omega}=\Delta\omega +\langle\xi\rangle +\langle B\rangle \epsilon$.
 Here $\langle A \rangle$  and $\langle B \rangle $ are the mean value  of $A$
 and $B$ respectively. This equation is similar to the one describing a phase
 locking of the periodic oscillator in the presence of noise \cite{Stratonovich}.

  In Eq. (4), if we turn off $\tilde{\xi}$, the analysis of the system
 stability is straightforward. If $\frac{d \phi}{dt} = 0$ the system ends
 time-evolution and $\phi$ remains at $\phi^*$, where
\begin{equation}
        \phi^* =\arcsin(-\frac{\Delta\tilde{\omega}}{\langle A \rangle}).
\end{equation} 
 Here the condition of $\phi^*$ being stable is
 $|\frac{\Delta\tilde{\omega}}{\langle A \rangle}| \leq 1$. Tangent bifurcation
 occurs at $\frac{\Delta\tilde{\omega}}{\langle A \rangle}=-1$ and the tangent
 point is $\phi^*_\pm=\pm\frac{\pi}{2}\pm 2\pi n$, where $n= 0,1,2,...$ and
 the sign $\pm$ depends on $\Delta \omega$. In our system since $\Delta \omega$
 is positive, only $\phi^*_+$ appears. This explains why $\phi$ is locked near
 $\pm\frac{\pi}{2}\pm 2\pi n$ and $2\pi$ phase jumps
 occur\cite{PhaseSync1,Lee}. 

\section{Numerical Study and return maps}

  If Eq. (4) is expanded around the tangent point $\phi^*_+$, the following
 equation is obtained:
 $\frac{d\tilde{\phi}}{dt}\approx\tilde{\epsilon}+a\tilde{\phi}^2+\tilde{\xi}$,
 where $\tilde{\phi}=\phi-\pi/2$,
 $\tilde{\epsilon}=\Delta\tilde{\omega}+\langle A\rangle$, and
 $a=-\langle A\rangle/2$. Here if $\tilde{\xi}$ is absent, $\tilde{\phi}$
 moves very slowly around the tangent point $\phi^*$. So the dynamics of
 $\tilde{\phi}$ is mainly governed by $\tilde{\xi}$ in the situation
 $|\tilde{\xi}| \gg |\tilde{\epsilon}|$. Then we can regard
 $\tilde{\phi}^2$ as a constant when the two oscillators are in a locked state.
 We obtain a local Poincar\'{e} map by integrating the above equation during
 the period that oscillator 1 completes every N rotation (the structure of the
 local Poincar\'{e} map is invariant with respect to the number of N as far
 as N is small enough in comparison with the average length of the phase locking
 state). The local Poincar\'{e} map is given by:
\begin{equation}
        \phi_{n+1}=\phi_n + \tilde{\epsilon}' + \tilde{a} \phi_n^2 + \xi_n,
\end{equation}
 where $\tilde{\epsilon}'=\tilde{\epsilon}\langle T_n \rangle$,
 $\tilde{a} = \frac{1}{2} \langle \int _{\tau_{n-1}}^{\tau_n} A dt \rangle$,
 and $\xi_n =\int_{\tau_{n-1}}^{\tau_n} \tilde{\xi} dt + \frac{1}{2}
 (\int _{\tau_{n-1}}^{\tau_n} A dt- \langle \int _{\tau_{n-1}}^{\tau_n} A dt \rangle)$.
 Here $T_n$ is $\tau_n - \tau_{n-1}$, where $\tau_n$ is the overall time that
 oscillator 1 takes to complete $N$ rotations. This is the very local
 Poincar\'e map of type-I intermittency in the presence of noise\cite{Hirsch}
 when $\xi_n$ acts as random noise. In this equation tangent bifurcation occurs
 at the point $\tilde{\epsilon}' = 0$, which meets the condition of
 $\frac{\Delta\tilde{\omega}}{\langle A \rangle} = -1$.
 To determine the tangent bifurcation point, we calculate $\langle A \rangle$,
 $\langle B \rangle$, $\langle \xi \rangle$,
 $\frac{\Delta\tilde{\omega}}{\langle A \rangle}$, and
 $\tilde{\epsilon}$ according to the coupling strength $\epsilon$ as presented
 in Fig. 1 (a), (b), and (c), respectively. The figures show that the value of
 $\epsilon$ at the bifurcation point is $0.0277$.

\begin{figure}
\begin{center}
\rotatebox[origin=c]{0}{\includegraphics[width=8.5cm]{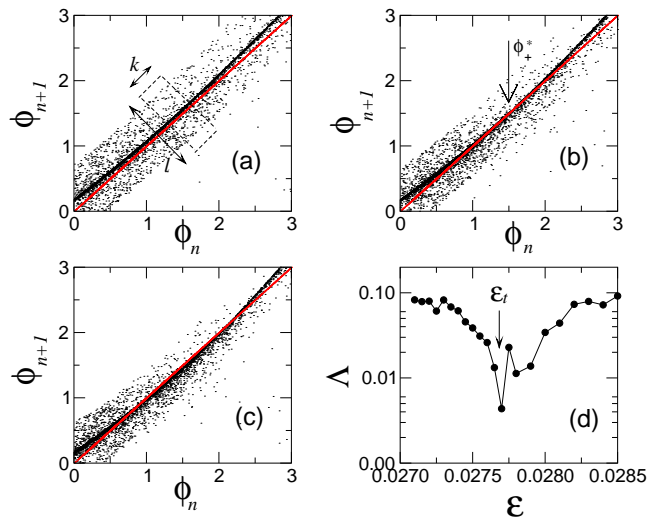}}
\caption{The return maps $(\phi_{n}, \phi_{n+1})$ and
   $N$ as a function of $\epsilon$: (a), (b), and (c)
   are the return maps before ($\epsilon = 0.018$), near ($\epsilon=0.028$),
   and after ($\epsilon=0.035$) the tangent bifurcation point, respectively;
  (d) shows that the minimum value of $\Lambda$ appears at $\epsilon_t=0.0277$ (when  $k=0.03$ and $l=0.6$).}
\end{center}
\end{figure}

  In order to verify the above analytic results, we construct the return maps
 directly from Eqs. (1)-(3) by obtaining $\phi$ for $N=1$. Figure 2(a), (b),
 and (c) show the return maps before, near, and after the tangent bifurcation
 point, respectively. The figures show that as the coupling strength $\epsilon$
 increases, the shadow curve approaches the diagonal line. When the curve
 touches the diagonal line, tangent bifurcation occurs and the tangent point
 is $\frac{\pi}{2}$ (see the inset arrow in Fig. 2(b)), which agrees well with
 the one obtained in the above.

  We define a measure $\Lambda =|(n_+ - n_-)/(n_+ + n_-)|$, where
 $n_+$ is the number of points above the diagonal line in the total 
number of points inside of the rectangle of $k \times l$ and
$n_-$ is the number of points below the diagonal line. 
Then the measure shows the average ratio between 
above and below the passage near the tangent point. 
So the minimum value of $\Lambda$
indicates the bifurcation point.
Figure 2(d) shows a sharp minimum at
 $\epsilon\approx 0.0277$. This value again agrees with what we obtained
 from Fig. 1(b) and (c).

  The shadow curves are well fitted to the following form of
 type-I intermittency\cite{PM,Kim}:
\begin{equation}
     \hat{\phi}_{n+1}=\hat{\phi}_n+ \hat{a} \hat{\phi}^{2}_n + \hat{\epsilon},
\end{equation}
 where $\hat{\phi}_n=\phi_n-\frac{\pi}{2}$, $\hat{a} \approx 0.094$, and
 $\hat{\epsilon}=\epsilon_t-\epsilon$. The coefficient $\hat{a}$ in Eq. (7)
 agrees well with $\tilde{a}$ in Eq. (6), since
 $\langle T_n \rangle \approx 2\pi$ and the mean value of $A$ is $-0.03$.
 (The overall frequency $\omega_{1,2} \approx 1.0$.) This confirms that the
 phase equation of the coupled R\"ossler oscillators coincide with the
 structure of type-I intermittency. 

\section{Result from Fokker-Planck Equation}

  Under the long laminar length approximation, Eq. (6) can be transformed into
 the differential form
 $\frac{d\phi}{dt} = \tilde{\epsilon}' + \tilde{a} {\phi}^2 +\tilde{\xi}$.
 Then we can obtain the Fokker-Planck equation (FPE) by regarding $\xi_n$ as
 Gaussian white noise\cite{Gardiner,FPE}. The probability distribution and
 auto-correlation of $\xi_n$ are examined for various $N$s. So we find they are in
 better agreement with the Gaussian distribution and $\delta$-correlation,
 respectively, as $N$ becomes larger. Figure 3 shows the probability
 distribution and auto-correlation of $\xi_n$ for $N=25$ which coincide 
 with the Gaussian profile with a dispersion of $0.4$ and the
 $\delta$-function, respectively. ($N=25$ is still much less than 
 the average length of the phase locking state that we have obtained.) The noise with the
 Gaussian distribution is not bounded whereas $\xi_n$ in coupled R\"{o}ssler
 oscillators is bounded within $(-1.3, 1.3)$. However, from the numerical
 data, we find that the probability of the occurrence of events outside the
 bounded region of $\xi_n$ is less than $1.0 \times 10^{-8}\%$. So it is
 a negligible effect on the average length of the phase locking state.

\begin{figure}
\begin{center}
\rotatebox[origin=c]{0}{\includegraphics[width=8.5cm]{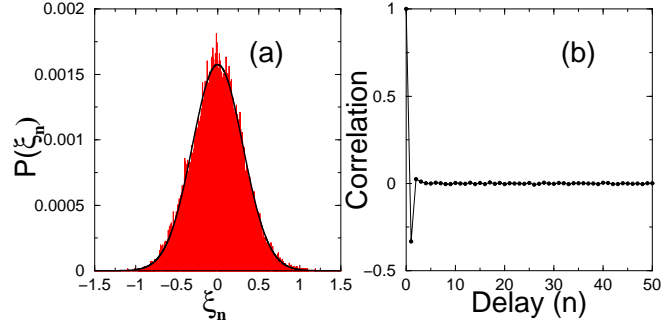}}
\end{center}
\caption{The distribution and auto-correlation of $\xi_n$ for $N=25$.
   (a) the distribution of $\xi_n$ well fitted to the Gaussian profile, and
   (b) the auto-correlation of $\xi_n$ almost $\delta$-correlated.}
\end{figure}

  From the FPE with appropriate boundary
 conditions\cite{Kye,Eckman,Gardiner,Hirsch},
 we can obtain the characteristic form of the average laminar length
 according to the coupling strength $\epsilon$ as follows:
\begin{equation}
        \langle l \rangle \cong \langle l_0 \rangle \exp( \alpha |\epsilon_t -\epsilon|^{3/2}), 
\end{equation}
 where $\alpha$ is the constant and $\langle l_0 \rangle$ is 
 the average length of the phase locking state 
 at the tangent bifurcation point. Figure 4(a) shows 
 the average length of the phase locking state 
 for $N=1$ as a function of $|\epsilon_t - \epsilon|$ in
 the region $0.0200 \leq \epsilon \leq 0.0302$. 
 The slope in the space $\ln|\epsilon_t -\epsilon|$ versus
 $\ln(\ln\langle l \rangle -\ln \langle l_0 \rangle)$ is $1.44$ as shown
 in Fig. 4(b) and its inset (c). The line fits well within $4.0$\% error from the 3/2 slope.
 The slope of the tail in Fig. 4 (b) converges to 1 \cite{Slope1} which shows the
 transient regime from $\langle l \rangle \propto (\epsilon_t -\epsilon)^{-1/2}$ to
 $\langle l \rangle \propto \exp(\alpha |\epsilon_t -\epsilon|^{3/2})$.  
 Figure 4(d) is the plot of $\ln(\epsilon_t -\epsilon)$ versus
 $\ln\langle l \rangle$ in the region $0.024 \leq \epsilon \leq 0.0277$ 
 and the tail also clearly shows the transient regime. 
The straight line fits well with the $-1/2$ slope. This means that the
 characteristic relation deforms from the conventional scaling rule
 $\langle l \rangle \propto (\epsilon_t -\epsilon)^{-1/2}$ to
 $\langle l \rangle \propto \exp( \alpha |\epsilon_t -\epsilon|^{3/2})$, as
 the coupling strength crosses the tangent bifurcation point. Thus we can
 understand that the average length of the phase locking state agrees well with the
 characteristic relation of type-I intermittency in the presence of noise not
 only in the region $\epsilon_t -\epsilon < 0$ but also
 in the region $\epsilon_t -\epsilon > 0$.

\begin{figure}
\begin{center}
\rotatebox[origin=c]{0}{\includegraphics[width=8.5cm]{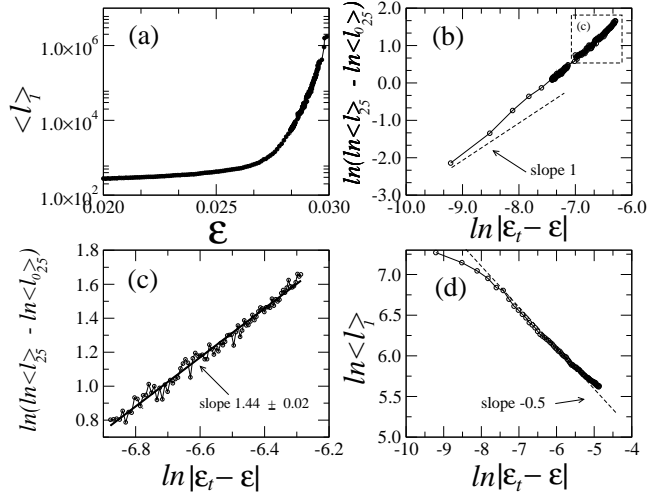}}
\caption{The numerical verification of the characteristic relation:
   (a) is $\langle l \rangle$ versus $\epsilon$ in the region
   $0.0200 \leq \epsilon \leq 0.0302$; (b) is
   $\ln(\ln\langle l \rangle_{25} -\ln\langle l_0 \rangle_{25})$ versus
   $\ln|\epsilon_t - \epsilon|$ after tangent bifurcation
   (where $\ln\langle l_0 \rangle_{25}=2.34\cdots$); (c) shows the magnified view of the inset in (b); 
   (d) is $\ln\langle l \rangle_1$ versus $\ln|\epsilon_t - \epsilon|$ before tangent
   bifurcation. The average length of the phase locking state is the average rotation numbers
   such that $\langle l \rangle_N = \langle T \rangle /2\pi N$, where
   $\langle T \rangle$ is the average phase locking time. }
\end{center}
\end{figure}

\section{Discussions}
  In the previous study, Lee {\it et al.}\cite{Lee} once showed that the characteristic
 relation of the average length of the phase locking state for $(\epsilon_t -\epsilon) >0$
 $(\epsilon_t -\epsilon)^{-1/2}$ that is the scaling of type-I intermittency.
 And then they claimed that the relation deforms to the scaling of eyelet
 intermittency for $(\epsilon_t -\epsilon)<0$ based on the numerical
 data only in the narrow range $0.0276 \leq \epsilon < 0.0286$. 
 They obtained the tangent bifurcation point
 $\epsilon_t=0.0276$ and the critical point for PS $\epsilon_c=0.0286$ 
 neglecting all the fast fluctuation terms, which are highly 
 important in this regard as we have explained above. 
Also a monograph claimed that the critical point for PS is $\epsilon_c \sim 0.028$ 
 relying based on the Lyapunov exponent analysis\cite{Appl}. 
 Unlike their claims, however, we have showed that the true tangent
 bifurcation point is $\epsilon_t = 0.0277$ and we have obtained the characteristic
 relation in the wider range $0.0200 \leq \epsilon \leq 0.0302$ where it
 deforms from $\langle l \rangle \propto (\epsilon_t -\epsilon)^{-1/2}$ to
 $\langle l \rangle \propto \exp(\alpha|\epsilon_t-\epsilon|^{3/2})$
 continuously as $\epsilon$ crosses $\epsilon_t = 0.0277$. This deformation is
 the typical characteristic of type-I intermittency in the presence of noise, as
 it was confirmed experimentally in our recent paper\cite{Jin}. In numerical
 simulations, we have also found that phase jumps still occur even at
 $\epsilon=0.0304$. As mentioned in the above, $\xi_n$ is bounded in the
 coupled R\"{o}ssler oscillators. Nevertheless, it is very hard to determine the
 correct $\epsilon_c$ because it takes too long time to find out the point
 where the average length of the phase locking state becomes infinite due to the exponential
 increment of $\langle l \rangle$. Instead of the PS point, the result of numerical
 simulation follows well Eq. (8) in the region which we studied.

In the route to PS, the phase slip phenomenon in coupled chaotic oscillators
is usually described by the Langevin equation: $\frac{d\phi}{dt} = -\frac{dV}{d\phi}+\tilde{\xi}$ \cite{Tut,PhaseSync1,Lee,Kye}.
We understand that the phenomenon is in a process of losing the stability for the fixed point
$\phi ^*$ where $-\frac{dV}{d\phi}|_{\phi^*}=0$ by the stochastic perturbation \cite{Kye,Gardiner,FPE}
and so the origin of the transition seems to be universal in coupled chaotic oscillators.
The characteristic scaling rule can be deformed according to the local structure
of the Poincar\'e map near the bifurcation point i.e.,  $-\frac{dV}{d\phi}$.
 Recently we observed a similar transition route which has the same origin in coupled
 hyper-chaotic R\"{o}ssler oscillators whose characteristic scaling rule 
 is governed by type-II intermittency
 in the presence of noise because its normal 
 form has cubic polynomial type instead of a quadratic one\cite{IBKim}.

\section{Conclusions}

  We have studied the origin of the transition to PS via phase jumps
 in coupled R\"{o}ssler oscillators analytically as well as numerically.
 Analysis of the phase equation and the numerically constructed local
 Poincar\'e map reveal that the transition to PS via $2\pi$ phase jumps is
 governed by type-I intermittency in the presence of external additive noise.
 The characteristic behavior of the average length of the phase locking state
 with respect to the parameter $\epsilon$ obtained both by numerical fitting
 and the FPE approach obeys $\langle l \rangle \cong \langle l_0 \rangle
 \exp(\alpha|\epsilon - \epsilon_t|^{3/2})$ for $\epsilon > \epsilon_t$, with
 the well known scaling form
 $\langle l \rangle \propto 1/\sqrt{|\epsilon-\epsilon_t|}$ for $\epsilon < \epsilon_t$.

\section{Acknowledgements}

 Authors thank S.-Y. Lee for helpful discussions.
  This work is supported by
 Creative Research Initiatives of the Korean Ministry of Science and
 Technology.


\begin{thebibliography}{150}
\bibitem{Slope1} The scaling rule near the tangent point can be estimated
as follows. Average laminar length is the function of
coupling strength such that $\langle l \rangle \sim f(|\epsilon_t-\epsilon|)$.
If $|\epsilon_t-\epsilon| \ll 1 $ then $f(|\epsilon_t-\epsilon|) \approx
f(0) + f^\prime(0) |\epsilon_t-\epsilon|+ O(|\epsilon_t-\epsilon |^2)$.
It means that  $f(|\epsilon_t-\epsilon|) \approx f(0) \exp(f^\prime(0)/f(0) |\epsilon_t -\epsilon|)$.
Thus average laminar length is given by 
$\langle l \rangle \sim \exp(\alpha |\epsilon_t-\epsilon|)$ for $|\epsilon_t-\epsilon| \ll 1 $,
where $\alpha$ is a constant.
\end{thebibliography}
\end{document}